\documentclass[aps,prl,twocolumn]{revtex4}

\usepackage{amsmath}
\usepackage{latexsym}
\usepackage{graphicx}
\usepackage{color}
\usepackage{bm}
\newcommand{\mm}{\,\mbox{mm}}

\newcommand{\muW}{\,\mu\mbox{W}}
\newcommand{\mW}{\,\mbox{mW}}
\newcommand{\nm}{\,\mbox{nm}}

\newcommand{\ns}{\,\mbox{ns}}

\newcommand{\fs}{\,\mbox{fs}}

\begin{document}

\title{Classical analog for dispersion cancellation of entangled photons with local detection}
\author{R. Prevedel}\email{robert.prevedel@iqc.ca}
\author{K.M. Schreiter}
\author{J. Lavoie}
\author{K.J. Resch}\email{kresch@iqc.ca}
\address{Institute for Quantum Computing and Department of
Physics \& Astronomy, University of Waterloo, Waterloo, Canada,
N2L 3G1}

\begin{abstract}
\noindent Energy-time entangled photon pairs remain tightly
correlated in time when the photons are passed through equal
magnitude, but opposite in sign, dispersion.  A recent
experimental demonstration has observed this effect on
ultrafast time-scales using second-harmonic generation of the
photon pairs.  However, the experimental signature of this
effect does not require energy-time entanglement. Here, we
demonstrate a directly analogue to this effect in narrow-band second
harmonic generation of a pair of classical laser pulses under
similar conditions.  Perfect cancellation is observed for fs
pulses with dispersion as large as $850$~fs$^2$, comparable to
the quantum result, but with an $10^{13}$-fold improvement in signal brightness.
\end{abstract}
%

\maketitle

\emph{Introduction} -- Ultrafast pulses of light are
indispensable tools at the heart of technologies as diverse as
time-resolved spectroscopy, fusion energy, surgery,
micro-machining, and medical imaging~\cite{diels06}. These
pulses require a very large bandwidth of light, all locked
together in phase, to support them. This makes the pulses very
susceptible to the effects of dispersion, the change of the
group velocity of the light with respect to frequency.
Dispersion slows some of the frequency components of the pulse
relative to others and spreads the pulse in time, an effect
that becomes more pronounced with larger bandwidth. If the
timing information carried by the pulse is critical to the
application, such as clock synchronization~\cite{giovannetti01}
or interferometry, this is clearly detrimental.

In 1992, two seminal papers \cite{steinberg92,franson92} showed
that energy-time entangled photons exhibit an inherent
robustness against dispersion, in two very different scenarios.
The first scenario considered the Hong-Ou-Mandel two-photon
interferometer with unbalanced dispersion, such as that from an extra
piece of glass, in one arm.  The Hong-Ou-Mandel interference
dip~\cite{Hong87}, whose width is equal to the coherence length of the
photons, is completely unaffected by all \emph{even} orders of
dispersion \cite{steinberg92b,resch09footnote,resch09}.  This
effect has been shown to have classical analogues
\cite{erkmen06,banaszek07,resch07,kaltenbaek08,legouet10}.

The second scenario was considered by Franson \cite{franson92}.
If a pair of transform-limited classical light pulses is sent
to different detectors, then any quadratic dispersion during
the propagation hurts their temporal correlation.  Recently it
was shown that the effect on the correlation for classical
pulses can be expressed as an inequality
\cite{wasak10},
\begin{eqnarray}
\label{inequality}
\langle \Delta \tau_F ^2\rangle \ge \langle \Delta \tau ^2\rangle + \frac{(2\beta)^2}{\langle \Delta \tau^2 \rangle}
\end{eqnarray}
where $\langle \Delta \tau_F ^2\rangle$ is the final variance
in the time difference of the detection signals, $\langle
\Delta \tau^2\rangle$ is the initial variance, and $\beta$
characterizes the dispersion which applies a quadratic
frequency dependent phase $\phi(\omega)=\beta(\omega-\omega_0)^2$
about some centre frequency $\omega_0$. Franson calculated that
if, instead, energy-time entangled photon pairs were sent to
the different detectors, as shown in Fig. 1(a), the photon
pairs remained tightly coincident in time when one photon
passed through positive dispersion and the other passed through
equal magnitude \emph{negative} dispersion \cite{franson92}.
Thus Eq.~\ref{inequality}, can be violated for
entangled photons.

In order to observe this effect, one needs to introduce
dispersion significant on the timescale of the detector
response.  Direct detection of dispersion cancellation using
single-photon counters and coincidence detection requires large
dispersion on the $\ns$ scale, as was achieved in
Refs.~\cite{brendel98,baek09}. A very recent experiment by
O'Donnell \cite{odonnell11}, probes dispersion cancellation
with femtosecond resolution using second-harmonic generation of
photon pairs followed by photon counting of the up-converted
beam as an ultrafast coincidence detector
\cite{dayan05,odonnell09}. (The short coincidence window
originates from the speed of the nonlinear process and persists
even when the detector is orders of magnitude slower).  The
second-harmonic generation signal originates from
simultaneously arriving photons, but by varying an optical
delay, $\tau$, one can measure temporal correlations with a
time offset \cite{odonnell09}. Unlike the prior experiments
which follow Franson's original proposal \cite{franson92} by
sending the photons to different detectors (as in Fig.~1(a)),
O'Donnell's experiment necessarily recombines the photons for the second-harmonic process (Fig 1(b)), i.e., it uses \emph{local} detection. It has been argued that dispersion
cancellation cannot be observed with only classical resources
in the original, \emph{nonlocal}, scenario
\cite{franson92,wasak10}, albeit with a fair amount of recent
discussion
\cite{torrescompany09,franson09,shapiro10,franson10}; however,
with local detection these arguments no longer apply.

One approach that has proven fruitful for developing classical
analogues of quantum technologies \cite{bennink02,ferri05,hemmer06,bentley04,torres11} is that of time reversal
\cite{resch07b,kaltenbaek08,kaltenbaek09,lavoie09}. Under this
transformation down-conversion, which is exclusively quantum
mechanical, can be replaced by second-harmonic generation,
which is a classical process. In the present work, we show that
the same signal observed in O'Donnell's experiment with
entangled photons can, in fact, be observed in a completely
classical experiment. Schematically, our setup is shown in
Fig.~1(c) where it can be viewed as the time-reverse of
Franson's original proposal, Fig.~1(a).  A pair of
time-correlated short laser pulses are sent through two
dispersive media.  The resulting pulses undergo fast
second-harmonic generation and photons of frequency $2\omega_0$
are detected using a standard single-photon counter.

\begin{figure}[t!]
 \begin{center}
  \includegraphics[width=1\columnwidth]{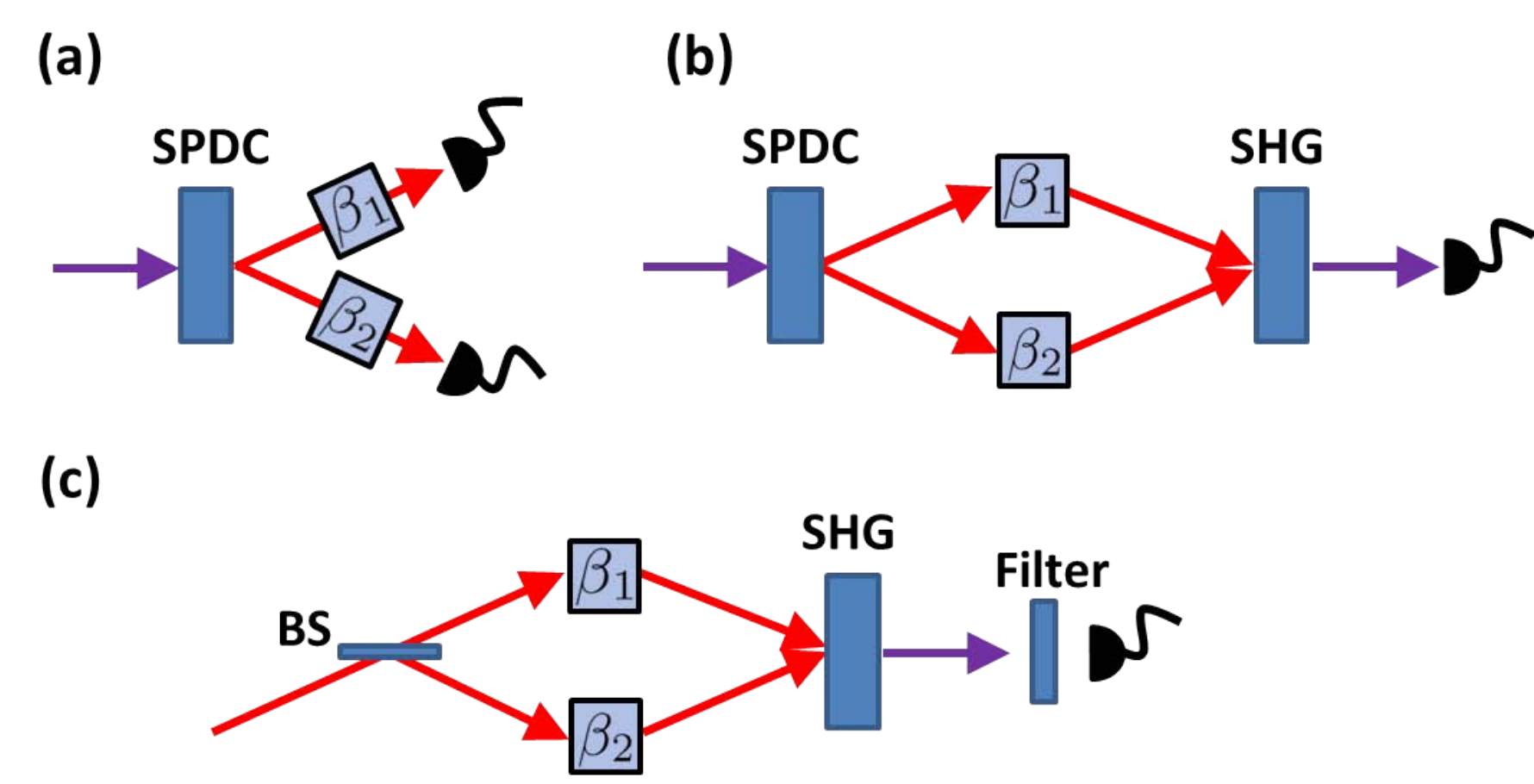}
 \end{center}
 \caption{(a) In Franson's proposal \cite{franson92}, a narrowband laser
 photon produces a pair of energy-time entangled
 photons via spontaneous parametric down-conversion (SPDC).  Each photon
 passes through different media characterized
 by second-order dispersion coefficient, $\beta$.
 Fast photon counters register the photon pairs and allow to
 determine the average time \emph{difference} in detection, $\langle \Delta \tau \rangle$. For perfectly energy-time entangled photons with $\beta_2=-\beta_1$, the effect of the dispersions cancel, and the pairs exhibit no additional broadening. (b) In O'Donnell's experiment \cite{odonnell11}, the photon counters
 and coincidence detection are replaced by second-harmonic generation which
 allows to probe dispersion effects on the $\fs$ timescale. (c) In the present
 work, we use a pair of classical laser pulses instead of entangled photons.
 We use second-harmonic generation, as in (b), but add narrow spectral filtering
 before detection of the second harmonic light.  The detected light has the same
 temporal characteristics of that in (b), including dispersion cancellation.}
 \label{fig1}
\end{figure}

\emph{Theory} --
In the literature, Franson dispersion cancellation has been
considered for the case of perfect energy-time entanglement. We
first reanalyze this effect using a physically-motivated model
with variable energy-time entanglement.  We consider the two-mode state,
\begin{eqnarray}
|\psi\rangle \sim \iint d\omega_1 d\omega_2 f(\omega_1,\omega_2)|\omega_1\rangle|\omega_2\rangle,
\end{eqnarray}
where we model the spectral function as in Ref.~\cite{resch09},
\begin{eqnarray}
f(\omega_1,\omega_2)=e^{-\frac{(\omega_1-\omega_0)^2}{2\sigma^2}}e^{-\frac{(\omega_2-\omega_0)^2}{2\sigma^2}}e^{-\frac{(\omega_1+\omega_2-2\omega0)^2}{2\sigma_c^2}}
\end{eqnarray}
with the spectrum of each photon given by a Gaussian
distribution centered at $\omega_0$ with rms width $\sigma$. Here, $\sigma_c$ describes the strength of the
energy correlation between the modes -- for down-conversion
sources, this parameter is given by the pump bandwidth. If
$\sigma_c \gg \sigma$, $f(\omega_1,\omega_2)$ factorizes and the
state is separable while if $\sigma_c \rightarrow 0$ then the
photons are perfectly energy anticorrelated and we obtain the
perfect energy-time entangled state. Using the same approach as
in \cite{franson92}, we calculate the probability of detecting
one photon at time, $t_1$, and another at time $t_2$, using the
correlation function,
\begin{eqnarray}
P \propto\langle \hat{E}_1^{-}(x_1,t_1)\hat{E}_2^{-}(x_2,t_2)\hat{E}_1^{+}(x_1,t_1)\hat{E}_2^{+}(x_2,t_2)\rangle,
\end{eqnarray}
where the positive frequency electric field operator is
$\hat{E}_i^{+}(x,t)=i\sum_\omega \left(\frac{\hbar\omega}{2
\varepsilon_0 V}\right)^{1/2}e^{i (k x-\omega
t)}\hat{a_i}(\omega)$ and
$\hat{E}_i^{-}=(\hat{E}_i^{+})^\dagger$ \cite{loudon83}. As in
Ref.~\cite{franson92} we assume that the bandwidth is small so
we can remove $\sqrt{\omega}$ dependence from the summation. We
characterize the quadratic dispersion of each photon using the
coefficient $\beta=\frac{1}{2}\frac{d^2k}{d\omega^2}L$, where
$L$ is the length of the dispersive region, so that
$\phi(\omega)=k(\omega)L=\beta(\omega-\omega_0)^2$. With this simplifying
assumption and the condition considered in
Ref.~\cite{franson92} that $\beta_2=-\beta_1$, the coincidence
probability is,
\begin{eqnarray}
P(t_1,t_2)\propto \exp\left[{-\frac{(t_1-t_2)^2\sigma^4 + (t_1^2+t_2^2)\sigma^2\sigma_c^2}{2\sigma^2+\sigma_c^2+4\beta_1^2\sigma^4\sigma_c^2}}\right].
\end{eqnarray}
From this distribution we calculate the variance in the time
difference of detections, $\Delta \tau_Q^2 =
\Delta(t_1-t_2)^2$,
\begin{eqnarray}
\label{quantumwidth}
\Delta \tau_Q^2=\frac{1}{\sigma^2}\left(1+\frac{4\beta_1^2\sigma^4\sigma_c^2}{2\sigma^2+\sigma_c^2}\right).
\end{eqnarray}
This allows us to recover key results of
Ref.~\cite{franson92} in different limits.  In the limit of
perfect energy-time entanglement, $\sigma_c\rightarrow 0$, we
have perfect dispersion cancellation since $\Delta
\tau_Q^2=\frac{1}{\sigma^2}$ and the dispersion does not reduce
the temporal correlation at all, clearly violating the
inequality in Eq.~\ref{inequality}. In the opposite limit where
there is no energy-time entanglement, $\sigma_c\rightarrow
\infty$, we have $\Delta
\tau_Q^2=\frac{1}{\sigma^2}(1+4\beta_1^2\sigma^4)$, i.e., the
photon temporal correlations are as susceptible to dispersion
as classical pulses (cf. Ref.~\cite[Eq. 20]{franson92}) and
cannot violate Eq.~\ref{inequality}.

Although infinitely narrow pump bandwidth SPDC, i.e, $\sigma_c
\rightarrow 0$, is unphysical, it is experimentally
straightforward to have a pump bandwidth orders of magnitude
narrower than that of the photons, $\sigma_c \ll \sigma$. In
this case,
\begin{eqnarray}
\Delta \tau_Q^2 \approx
\frac{1}{\sigma^2}\left(1+2\beta_1^2\sigma^2\sigma_c^2\right).
\end{eqnarray}
Comparing this to the completely separable case we see that the
effective bandwidth $\sigma$ is lowered approximately to the
geometric average of the pump and photon bandwidth,
$\sqrt{\sigma\sigma_c}$.  Since this effective bandwidth is
much smaller, the effect of dispersion on the temporal
correlation is significantly reduced.

We now compare this quantum mechanical situation to the
classical scenario shown in Fig.~1(c). We model a
transform-limited classical pulse using its complex electric
field amplitude,
\begin{eqnarray}
E(\omega)\propto e^{-\frac{(\omega-\omega_0)^2}{2\sigma^2}}.
\end{eqnarray}
We approximate the SHG of a fast nonresonant three-wave mixing
interaction as,
\begin{eqnarray}
E_{SHG}(\omega)\propto\int E_1(\omega')E_2(\omega-\omega')d\omega',
\end{eqnarray}
which holds for a fast nonlinearity within a relatively narrow
frequency band (the latter approximation is similar to removing
the $\sqrt{\omega}$ outside the summation in the electric field
operator). Including the effect of dispersion in both arms and
a time delay, $\tau$, on one of the pulses, the second-harmonic
amplitude at frequency $\omega$ is,
\begin{eqnarray}
E_{SHG}(\omega)&\propto&\\
&& \hspace{-2cm}\int E_1(\omega')e^{i\beta_1(\omega'-\omega_0)^2}e^{-i\omega'\tau}E_2(\omega-\omega')e^{i\beta_2(\omega-\omega'-\omega_0)^2}d\omega'. \nonumber
\end{eqnarray}
The intensity of the light is then measured at a single
frequency using a monochromator with a finite frequency
resolution, $\sigma_s$. We model the monochromator response
using the resolution function
$S(\omega)=e^{-\frac{(\omega-2\omega_0)}{2\sigma_s^2}}$, so
that the detected signal is
\begin{eqnarray}
I \propto \int S(\omega)I(\omega) d\omega.
\end{eqnarray}
We calculate the expected signal, under the condition
$\beta_2=-\beta_1$, as a function of the time delay, $\tau$, to
be
\begin{eqnarray}
I(\tau)\propto \exp\left[-\frac{\sigma^2(\sigma^2+\sigma_s^2)\tau^2}{2(\sigma^2+\sigma_s^2+4\beta_1^2\sigma^4\sigma_s^2)}\right].
\end{eqnarray}
From this expression we can compute the variance,
\begin{eqnarray}
\label{classicalwidth}
\Delta \tau_C^2=\frac{1}{\sigma^2}\left(1+\frac{4\beta_1^2\sigma^4\sigma_s^2}{\sigma^2+\sigma_s^2}\right).
\end{eqnarray}
Our expressions for the variances of quantum and classical
signals, Eqs.~\ref{quantumwidth} and \ref{classicalwidth}, are
remarkably similar.  Both signals show perfect dispersion
cancellation in the limits of $\sigma_c$ or
$\sigma_s\rightarrow0$.  Thus the pump bandwidth in the quantum
case plays the same role as spectrometer resolution in the
classical case.  These will, in practice, put constraints on
the amount of dispersion cancellation possible.  Assuming
$\sigma_{c(s)}\ll\sigma$, then the condition $\beta
\sigma_{c(s)}\sigma<1$ is still required for significant
dispersion cancellation.

\begin{figure}[h!]
 \begin{center}
  \includegraphics[width=1\columnwidth]{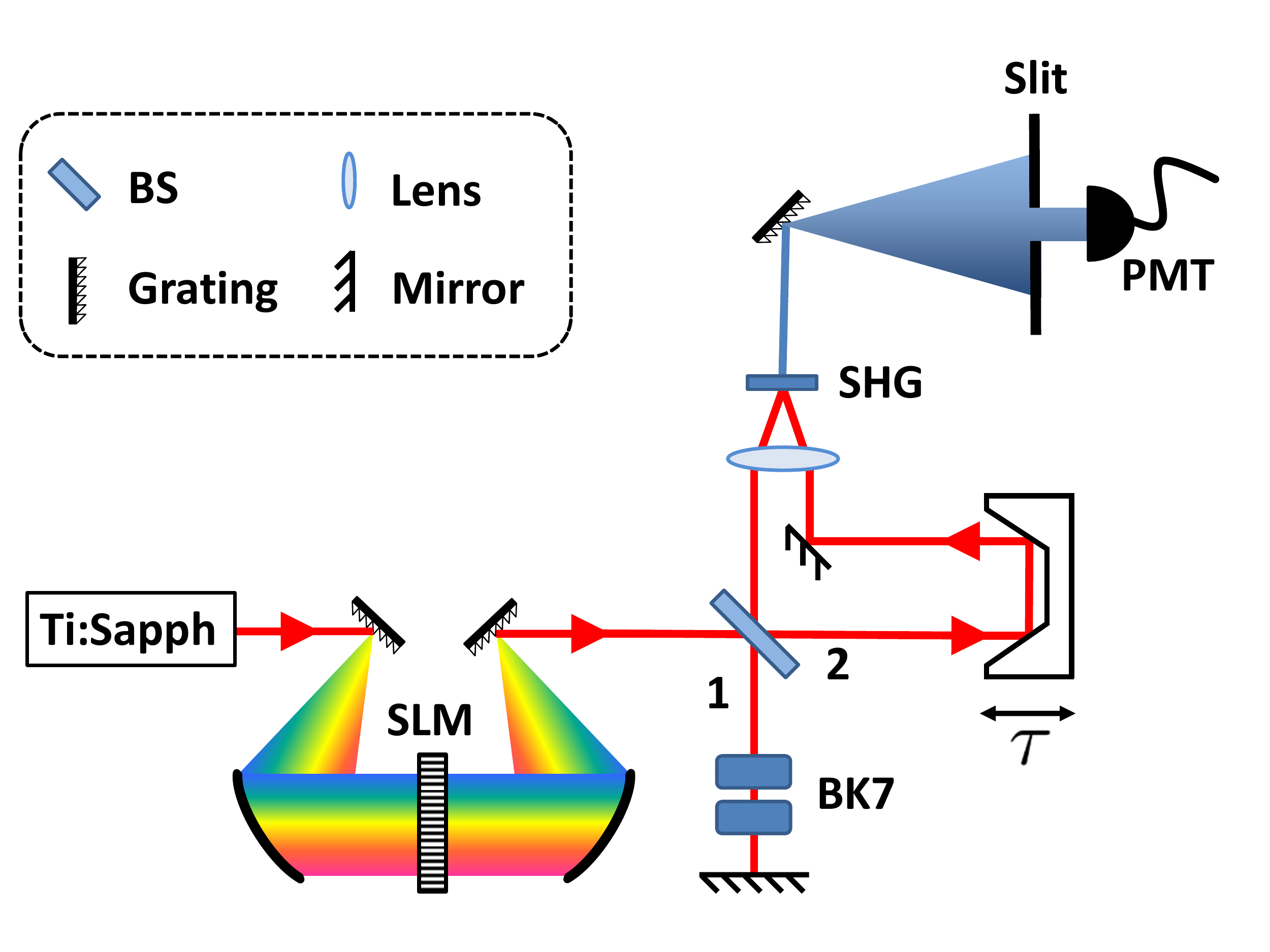}
 \end{center}
 \caption{Experimental setup. The pulsed laser light from the Ti:Sapph is sent through a 4-F pulse shaper consisting of two 1200 l/mm gratings,
 two curved mirrors and a 640 pixel spatial light modulator (SLM).
 Subsequently, the beams are separated into two arms by a 50/50
 beamsplitter (BS). In our experiment, positive dispersion in the sample arm (mode 1) can be introduced using two pieces of
 BK7 glass ($19.33\pm0.03\mm$ each), while the reference
 arm features an adjustable path delay, $\tau$. To observe dispersion
 cancellation, the beams are focused by a 75$\mm$ achromatic lens onto
 a 2$\mm$ BiBO nonlinear crystal and the upconverted light is
 spectrally filtered by a grating and a slit before detection by a PMT.}\label{setup}
\end{figure}

\emph{Experiment}--Our experimental setup is depicted in
Fig.~\ref{setup}.  Light from a pulsed titanium:sapphire laser
(KMLabs Griffin-10, centre wavelength
$807\nm$), passes through a 4-F pulse shaping apparatus
containing a spatial light modulator (SLM) (CRi 640 pixel, dual
mask) \cite{weiner00}.  The pulse shaper is used to recompress
the pulses from the laser and also to apply the negative
quadratic dispersion.  After exiting the pulse shaper, the
light is split into two paths using a 50/50 non-polarizing,
low-dispersion beamsplitter. We can introduce positive
dispersion into one of the arms using BK7 glass plates.  The
two beams are then redirected and focused together onto a 2-mm
thick bismuth borate (BiBO) crystal cut for second-harmonic
generation. The upconverted light, which had an average power of $30\mW$ when no additional dispersion was introduced,
is filtered from the remaining fundamental, passed through a monochromator (at wavelength $404\nm$) with 0.02$\nm$ FWHM resolution, and its
intensity measured by a photomultiplier tube (PMT). Using this
monochromator resolution and the bandwidth of the laser, we
expect that dispersion cancellation should persist for
dispersion up to
$\beta<1/\sigma_{c(s)}\sigma\approx7\times10^4\fs^2$ according
to the criterion derived in the previous section.

\begin{figure}[t!]
 \begin{center}
  \includegraphics[width=1\columnwidth]{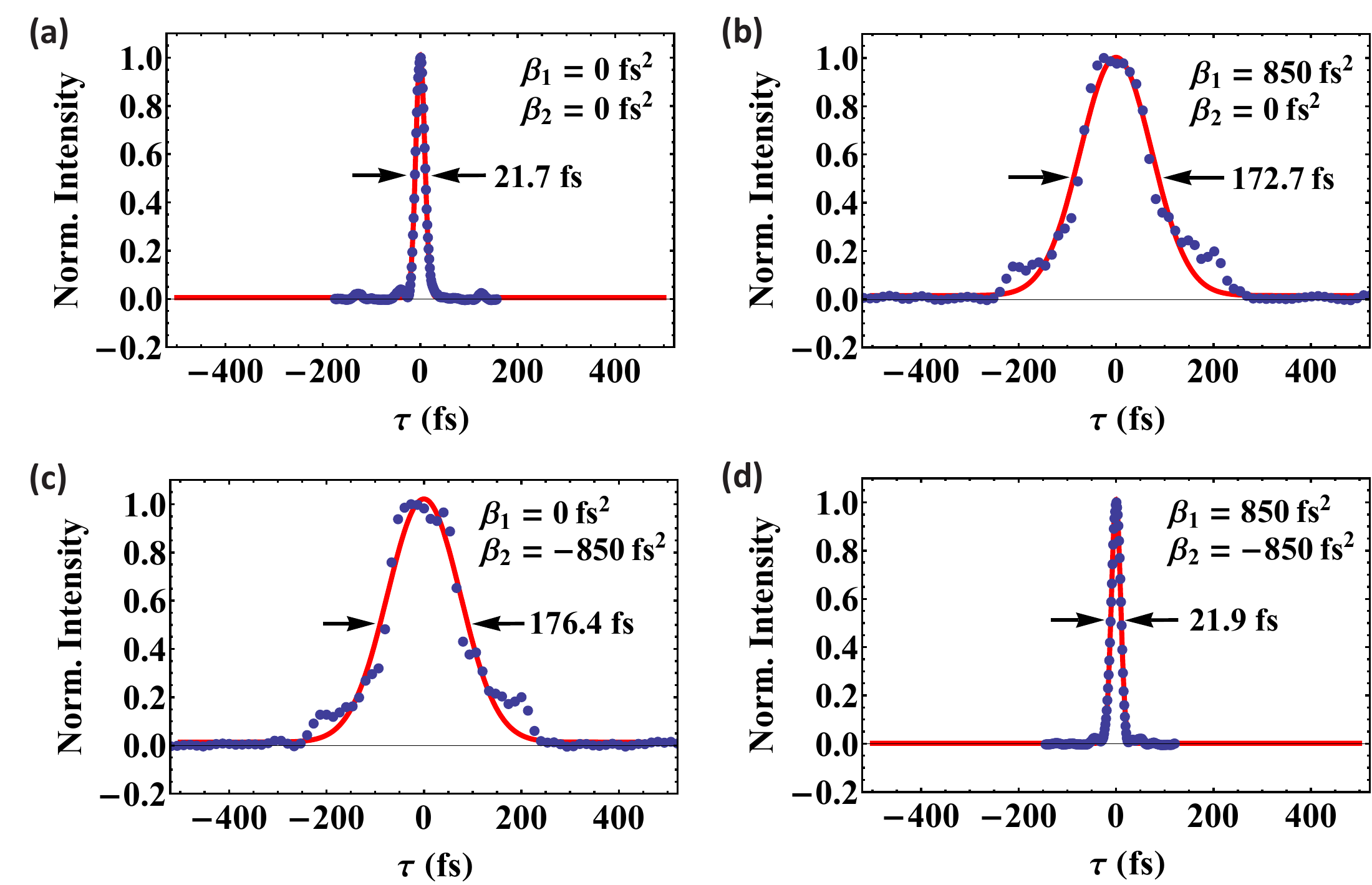}
 \end{center}
 \caption{Experimental data. The normalized intensity
 of the upconverted light is shown as a function of the temporal delay, $\Delta\tau$.
 (a) shows the dispersion-free case with a nearly Fourier-limited signal width
 ($21.7$fs). This is in stark contrast to the cases where
 dispersion of (b) $\beta = 850$~fs$^2$ is added to arm 1 or (c) $\beta = -850$~fs$^2$ is added to arm 2, where the
 signals broaden significantly to 172.7~fs and 176.4~fs, respectively.  However, adding positive dispersion
 of $\beta = 850$~fs$^2$ to arm 1 and negative dispersion $\beta = -850$~fs$^2$ to arm 2 yields the signal
 shown in (d) with a measured width of $21.9$~\fs. In this last case, we show the dispersion is cancelled in direct
 analogy with the quantum effect.}\label{data}
\end{figure}

We first compensated quadratic and cubic dispersion using the
pulse shaper to create a transform-limited laser pulse. Note that at this point there is no dispersion introduced in arm 1 of the interferometer. We
recorded the second-harmonic intensity through our
monochromator as a function of delay position, as shown in
Fig.~\ref{data}(a).  The data show a sharply peaked signal with
width $21.7\fs$ FWHM. This is in good agreement with
the estimated signal width of $20\fs$ for the measured FWHM bandwidth of our laser's electric field spectrum ($\Delta\lambda=97\nm$). We measured the amount of positive dispersion introduced by two identical pieces of BK7 glass, totalling $38.65\pm0.06\mm$
thickness, by inserting it in the path before the beamsplitter, and
applying negative quadratic dispersion with the SLM until we
achieved the same width in our SHG signal as in
Fig.~\ref{data}(a).  The dispersion of the glass was thus
measured to be $\beta=850\fs^2$ in good agreement with the theoretical value of $\beta=851\pm1\fs^2$, calculated from Sellmeier coefficients.

Using the transform-limited pulses, we inserted one piece of
glass into the arm 1 of the interferometer.  The light in that
arm makes two passes through the glass and thus experiences
$850\fs^2$ of dispersion.  The second-harmonic intensity as a
function of delay is shown in Fig.~\ref{data}(b), where the
signal is significantly broadened to $172.7$~fs.  Next, we used
the SLM to apply $-850\fs^2$ of dispersion, which compensates
the glass in arm 1, but leaves the beam in arm 2 with negative
dispersion. The second-harmonic signal in this case is shown in
Fig.~\ref{data}(c) and shows a similar level of broadening to
$176.4$~fs.  As expected here, and in agreement with the
quantum theory and experiment \cite{franson92,odonnell11},
dispersion in either arm leads to broadening of the temporal
correlation.

Finally, leaving the SLM applying $-850\fs^2$ we added a second
piece of glass to arm 1, so that the cumulative effect of the
SLM and two pieces of glass resulted in $+850\fs^2$ of net
dispersion for that arm, while the beam in arm 2 remained at
$-850\fs^2$. The second-harmonic data for this case is shown in
Fig.~\ref{data}(d). The measured width is $21.9\fs^2$,
unchanged from the zero dispersion case.  This is our classical
analogue to dispersion cancellation with local detection.
Our measured signal of $750\muW$ corresponds to a photon flux of $6\times10^{16}s^{-1}$, an $10^{13}$ increase over the signal reported in Ref.~\cite{odonnell11}.

\emph{Conclusion} -- Franson dispersion cancellation reveals a curious
robustness of temporal correlations in energy-time
entanglement.  However, once local detection is introduced, it
is straightforward to observe dispersion cancellation in a
classical experiment, with signals 13 orders of magnitude higher than in a recent state of the art quantum experiment~\cite{odonnell11}. It is interesting to consider whether applications in interferometry or clock synchronization could make use of the effect with the local detection constraint, as their
performance would dramatically improve from the increase in
signal alone. Regardless, our understanding of both theories is
improved by exploring the limits of classical physics to
emulate quantum mechanical effects.

\emph{Acknowledgements} -- We thank K. Shalm and R. Kaltenbaek for valuable discussions and are grateful for financial support from Ontario Ministry of Research and Innovation ERA, QuantumWorks, NSERC, OCE, Industry Canada and CFI. R.P. acknowledges support by MRI and the Austrian Science Fund (FWF).


\begin{thebibliography}{99}
\bibitem{resch09footnote} Even-order dispersion
    cancellation in the Hong-Ou-Mandel interferometer is
    ideal only when the energy-time entanglement is perfect.
    Once finite correlations are considered, the degree of cancellation
    is limited. However, with reasonable experimental
    parameters the impact on the interferometer signal under
    considerable amounts of dispersion can be dramatically reduced~\cite{resch09}.
\end{thebibliography}
\end{document}